\documentclass[journal]{IEEEtran}
\usepackage{graphicx,here}

\begin{document}
\title{Test beam results of the GE1/1 prototype for a 
future upgrade of the CMS high-$\eta$ muon system\\
\vspace*{-4cm}{\tiny This work has been submitted to the IEEE
  Nucl. Sci. Symp. 2011 for publication in the conference record. Copyright
  may be transferred without notice, after which this version may no longer be
  available.}\\
\hspace*{15cm}{\bf\small RD51-Note-2011-013}\vspace*{2cm}
}

\author{
D.~Abbaneo, M.~Abbrescia, C.~Armagnaud, P.~Aspell, M.~G.~Bagliesi, Y.~Ban, S.~Bally, L.~Benussi, U.~Berzano, S.~Bianco, J.~Bos, K.~Bunkowski,  J.~Cai, R.~Cecchi, J.~P.~Chatelain, J.~Christiansen, S.~Colafranceschi, A.~Colaleo, A.~Conde~Garcia, E.~David, G.~de~Robertis, R.~De~Oliveira, S.~Duarte Pinto, S.~Ferry, F.~Formenti, L.~Franconi, K.~Gnanvo, A.~Gutierrez, M.~Hohlmann~\IEEEmembership{Member,~IEEE}, P.~E.~Karchin, F.~Loddo, G.~Magazz\`u, M.~Maggi, A.~Marchioro, A.~Marinov, K.~Mehta, ,J.~Merlin, A.~Mohapatra, T.~Moulik, M.~V.~Nemallapudi, S.~Nuzzo, E.~Oliveri, D.~Piccolo, H.~Postema, G.~Raffone, A.~Rodrigues, L.~Ropelewski, G.~Saviano, A.~Sharma~\IEEEmembership{Senior Member,~IEEE}, M.~J.~Staib, H.~Teng, M.~Tytgat~\IEEEmembership{Member,~IEEE}, S.~A.~Tupputi, N.~Turini, N.~Smilkjovic, M.~Villa, N.~Zaganidis, M.~Zientek.

\thanks{Manuscript received November 15, 2011}

\thanks{M.~Abbrescia, A.~Colaleo, G.~de~Robertis, F.~Loddo, M.~Maggi, S.~Nuzzo, S.~A.~Tupputi are with Politecnico di Bari, Universit\`a di Bari, and INFN Sezione di Bari - Bari, Italy}
\thanks{Y.~Ban, J.~Cai, H.~Teng are with Peking University - Beijing, China}
\thanks{A.~Mohapatra, T.~Moulik are with NISER - Bhubaneswar, India}
\thanks{A.~Gutierrez, P.~E.~Karchin are with Wayne State University - Detroit, USA}
\thanks{L.~Benussi, S.~Bianco, S.~Colafranceschi, D.~Piccolo, G.~Raffone, G.~Saviano are with Labortori Nazionali di Frascati INFN - Frascati, Italy}
\thanks{D.~Abbaneo, C.~Armagnaud, P.~Aspell, S.~Bally, U.~Berzano, J.~Bos, K.~Bunkowski, J.~P.~Chatelain, J.~Christiansen, A.~Conde~Garcia, E.~David, R.~De~Oliveira, S.~Duarte Pinto, S.~Ferry, F.~Formenti, L.~Franconi, A.~Marchioro, K.~Mehta, J.~Merlin, M.~V.~Nemallapudi, H.~Postema, A.~Rodrigues, L.~Ropelewski, A.~Sharma, N.~Smilkjovic, M.~Villa, M.~Zientek are with Physics~Department,~CERN - Geneva,~Switzerland}
\thanks{A.~Marinov, M.~Tytgat, N.~Zaganidis are with Department of Physics and Astronomy Universiteit Gent - Gent, Belgium}
\thanks{K.~Gnanvo, M.~Hohlmann, M.~J.~Staib are with Florida Institute of Technology - Melbourne, USA}
\thanks{M.~G.~Bagliesi, R.~Cecchi, G.~Magazz\`u, E.~Olivieri, N.~Turini are with INFN Sezione di Pisa - Pisa, Italy}
\thanks{T.~Fruboes, is with Warsaw University - Warsaw, Poland}
\thanks{* Corresponding author: stefano.colafranceschi@cern.ch}
}

\maketitle
\pagestyle{empty}
\thispagestyle{empty}

\begin{abstract}
Gas Electron Multipliers (GEM) are an interesting technology under consideration for the future upgrade of the forward region of the CMS muon system, specifically in the $1.6<| \eta |<2.4$ endcap region.
With a sufficiently fine segmentation GEMs can provide precision tracking as well as fast trigger information. The main objective is to contribute to the improvement of the CMS muon trigger. The construction of large-area GEM detectors is challenging both from the technological and production aspects. In view of the CMS upgrade we have designed and built the largest full-size Triple-GEM muon detector, which is able to meet the stringent requirements given the hostile environment at the high-luminosity LHC.
Measurements were performed during several test beam campaigns at the CERN SPS in 2010 and 2011. The main issues under study are efficiency, spatial resolution and timing performance with different inter-electrode gap configurations and gas mixtures. In this paper results of the performance of the prototypes at the beam tests will be discussed.
\end{abstract}


\section{Introduction}
\IEEEPARstart{T}{}he GEM collaboration (GEMs for CMS) is developing a new prototype detector in view of the CMS high-$\eta$ upgrade\cite{tytgat_ieee2011}. The CMS\cite{:2008zzk} endcap region $1.6<| \eta |<2.4$ is not instrumented with the Resistive Plate Chambers\cite{Santonico:1994dk} (RPC) originally planned for this region. Gas Electron Multipliers\cite{Sauli:1997qp} (GEMs) are an interesting technology under consideration for a re-scope of this region of the CMS forward muon system.
The collaboration performed already several feasibility studies on small\cite{tytgat_ieee} and full-size\cite{colafranceschi_ieee} detectors, showing that GEMs are an appealing technology compared to the RPCs. While RPCs are fast, they cannot stand the strong and hostile environment, which is the reason why the high-$\eta$ region is still vacant. 
GEMs\cite{Alfonsi:2004jm} have been demonstrated to be a very mature technology with good spatial and time resolution, excellent high rate capability and radiation hardness. They can provide triggering and tracking functions at the same time. 
\par
In 2011, the collaboration successfully developed and tested the GE1/1$\_$II, which is the second full-scale prototype chamber, with enhanced performance relative to the GE1/1$\_$I\cite{colafranceschi_icatp11}, developed in 2010.

\section{Prototype description}
\begin{figure}[H]
  \centering
  \includegraphics[width=2.75in]{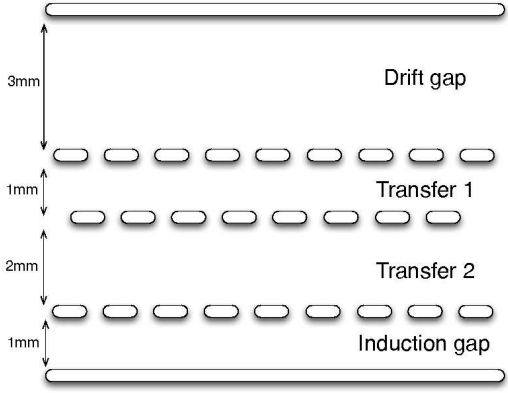}
  \caption{Gap configuration for the full-size prototype GE1/1$\_$II.}
  \label{gap_configuration}
\end{figure}

The GE1/1$\_$II represents the state-of-the-art of large-area GEM detectors. This prototype chamber hosts a Triple-GEM detector using a gap configuration  (shown in Fig.~\ref{gap_configuration}) with the following sizes: 3/1/2/1mm (drift,  transfer 1, transfer 2, induction gap sizes). This new gap configuration required an adjustment of the HV divider in order to provide optimal fields and voltages to the GEM foils.
This kind of configuration with small gaps is quite challenging since there are technological issues during the GEM foil stretching process and the assembly procedure. We have been working on optimizing the thermal stretching process in order to reduce possible foil ripples. Each foil was thermally stretched using a special oven with temperature of 37${^\circ}\rm{C}$ for 24 hours; since the transfer and induction gaps were just 1mm, the GEM foil stretching was extremely challenging because no ripple could be allowed. 
Once the stretching process is completed, the foil is glued together with the frame. The last step is the glue curing, again performed using the same oven. In every step of the process, GEM foils are tested with careful optical inspections and sector-by-sector HV tests in a dedicated clean room.

The external envelope of the chamber is a trapezoid with dimensions $990\rm{mm} \times (220-455)\rm{mm}$, as imposed by the CMS muon endcap high-$\eta$ area. The GE1/1$\_$II drift electrode is part of the aluminum chamber envelope itself and it is produced by gluing a $300 \rm{\mu m}$ kapton layer with $5 \rm{\mu m}$ copper cladding to a $3\rm{mm}$ aluminum plate. The GEM foil production relies on the novel photolithographic processes using a single-sided mask\cite{villa} developed at CERN applied to a $50 {\rm {\mu m}}$ thick kapton sheet with $5 {\rm {\mu m}}$ copper clad on both sides. This single-mask technique is used to overcome the problems with alignment of double masks which become critical once a GEM foil dimension exceeds $40\rm{cm}$.
The GEM foils are sectorized into 35 high voltage sectors transverse to the strip direction, where each sector has a surface area of about $\rm{100 cm^2}$ to limit the discharge probability. \\
The detector is divided into  8 $\eta$ partitions containing 384 readout strips each oriented radially along the long side of the detector. Every partition in $\eta$ is read out by three connectors for the electronics. 
The readout plane has been designed with 3184 channels using a variable strip pitch from 0.6mm at the narrow chamber end to 1.2mm at the wide end. 

For the strip readout, the binary VFAT2\cite{Aspell:2008zz} chip originally developed for TOTEM\cite{totem} and the analog APV25\cite{apv_chip} chip were adopted. 

The VFAT2 is a digital chip with 128 channels and it was designed with the $\rm{0.25\mu m}$ CMOS technology at CERN using radiation tolerant components. The chip offers a programmable fast OR function of the input channels and adjustable threshold, gain, and signal polarity, plus a programmable integration time of the analog input signal. The signal sampling of the VFAT2 chip is driven by a 40 MHz internal clock; during 2011 test beams the Monostable pulse length (MSPL) was set from 1 to 3 clock cycles. The VFAT2 has also an adjustable latency that is used to account for differences in delay due to the spatial spread of different detectors addressing specific properties of each detector in terms of signal shape and timing. The current of the comparator stage (IC) can be also set and experimentally it was found that low values drastically reduce the noise level. 

As data acquisition systems the TURBO system was used for the VFAT2 chip and the Scalable Readout System (SRS) recently developed by the RD51 collaboration\cite{rd51_collaboration} was used for the APV25 chip. The SRS will be used next year for testing new and old prototypes in the RD51 - CMS beam tests with both chips.

\section{Test Beam setups}
\begin{figure}[H]
  \centering
  \includegraphics[width=2.75in]{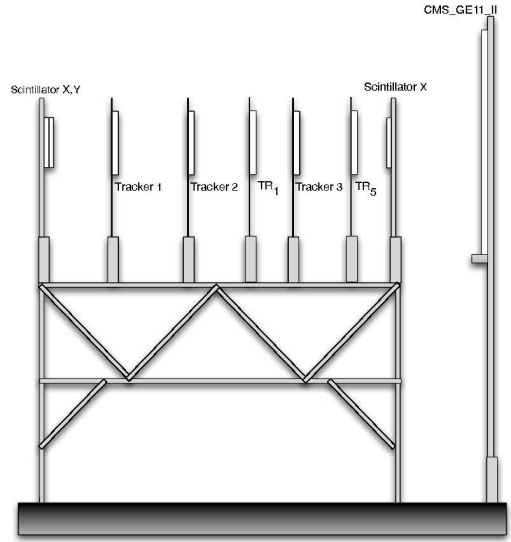}
  \caption{The RD51 telescope used during all test beams at the RD51 H4, H6, and H8 beam area.}
  \label{rd51_tracker_table}
\end{figure}

The full-scale detector GE1/1$\_$II was tested with a 150 GeV muon/pion beam at the CERN SPS H4, H6, and H8 beam lines during several RD51 test beam campaigns. The RD51 standard double-mask Triple-GEM beam telescope was used, as
depicted in Fig.~\ref{rd51_tracker_table}, as a tracking device for the GE1/1. 
The telescope consists of three standard Triple-GEM muon detectors, with a $\rm{10 \times 10 cm^2}$ active area, running with a $\rm{Ar/CO_2}$ (70:30) gas mixture and with a gain around $\rm{10^4}$.
Each detector of the telescope is equipped with 256 strips in both horizontal (y-coordinate) and vertical (x-coordinate) directions transverse to the beam, with a pitch of 0.4mm. 
The GE1/1$\_$II was installed near the RD51 telescope on a vertically movable table for scanning different points since the pitch of the full-scale detector is not constant.

\begin{figure}[H]
  \centering
  \includegraphics[width=1.71in]{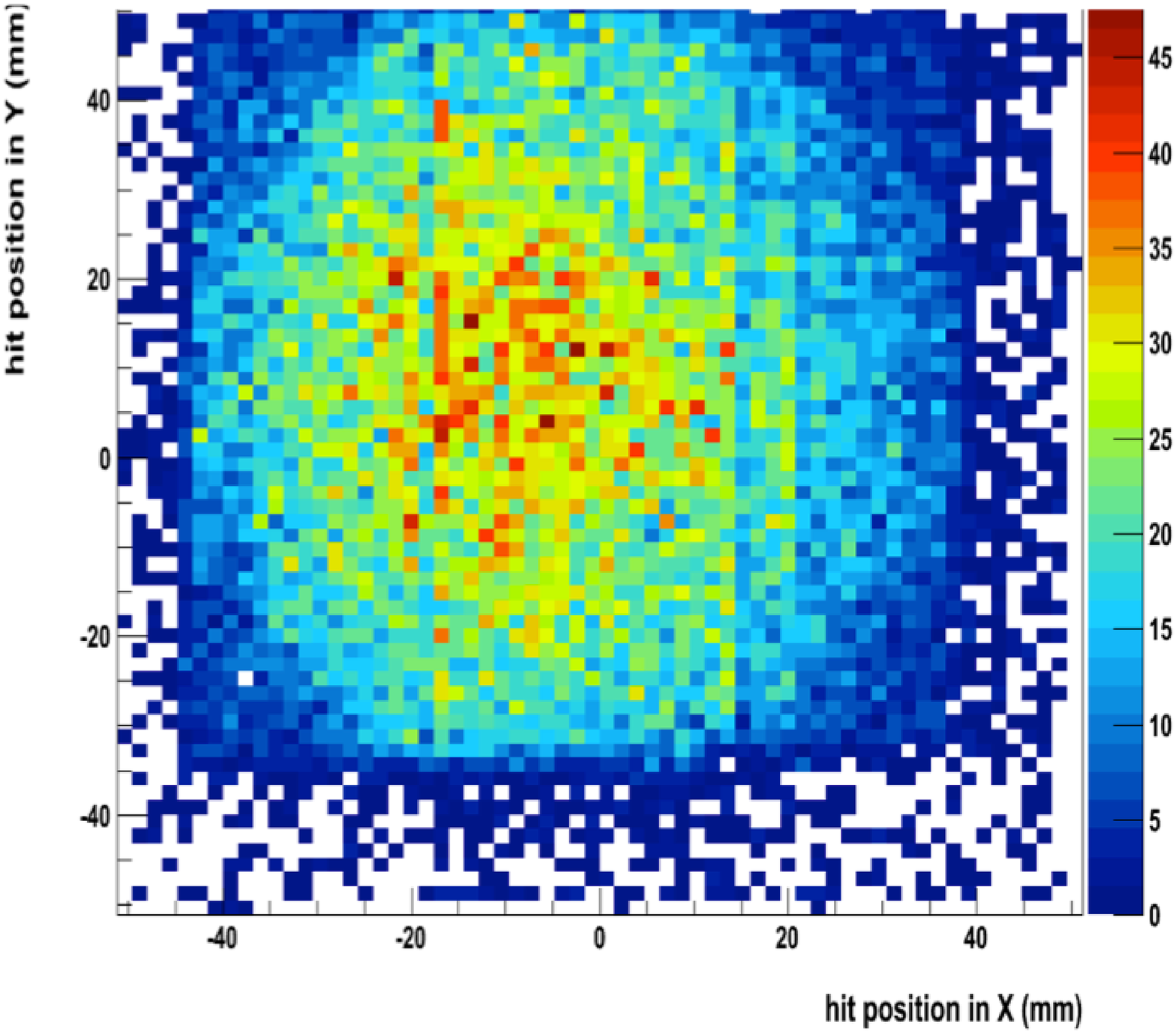}
  \includegraphics[width=1.71in]{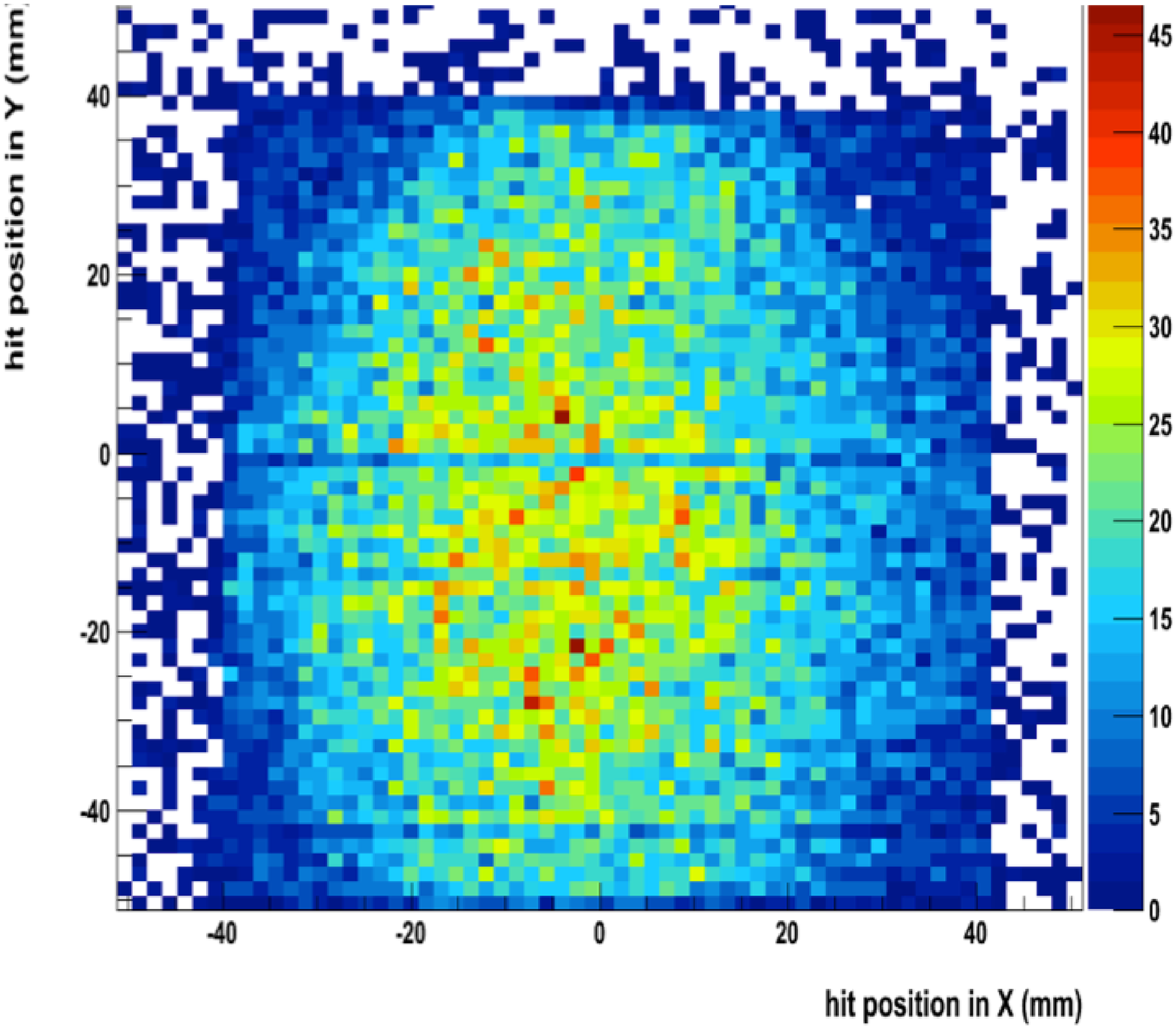}
  \caption{Typical tracker beam profile.}
  \label{beam_profile2011}
\end{figure}

Fig.~\ref{beam_profile2011} shows a typical beam profile of the muon beam as reconstructed with the GEM telescope. The reconstruction algorithm, after the alignment step, fits found hits on the telescope in both dimensions providing a space resolution of around $\rm{50\mu m}$.

\vspace{0.3cm}

In one test beam the CMS M1 superconductive magnet was operated to test the GE1/1$\_$II in a strong magnetic field. The RD51 Triple-GEM telescope as well as trigger PMTs were kept 5m away in order to be undisturbed. The M1 magnet is a solenoid that can achieve a field of 3T at $\approx$4000A; at this point the overall trigger DAQ efficiency is reduced since PMTs suffer from the fringe field.

\section{Results}
\begin{figure}[H]
  \centering
  \includegraphics[width=2.85in]{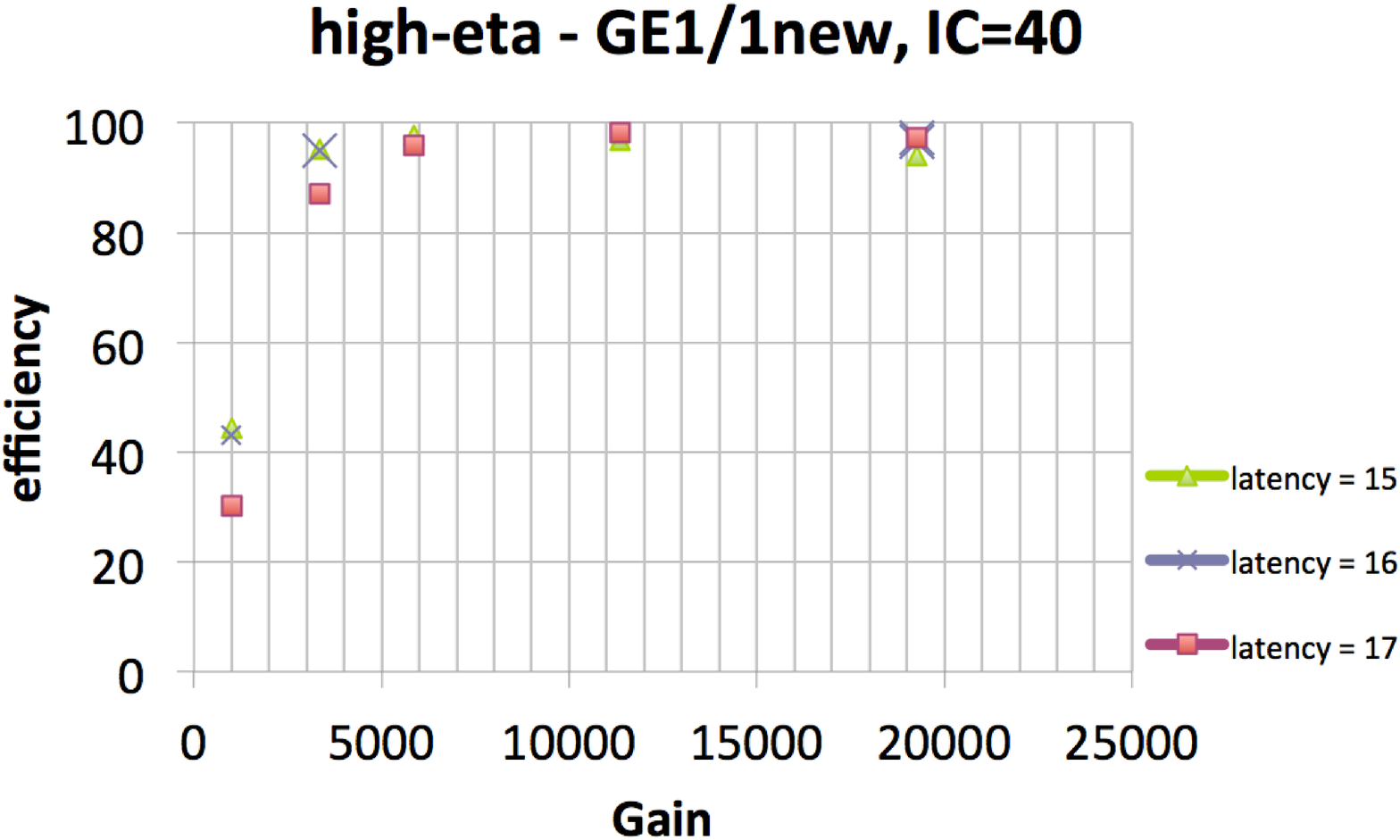}
  \includegraphics[width=2.85in]{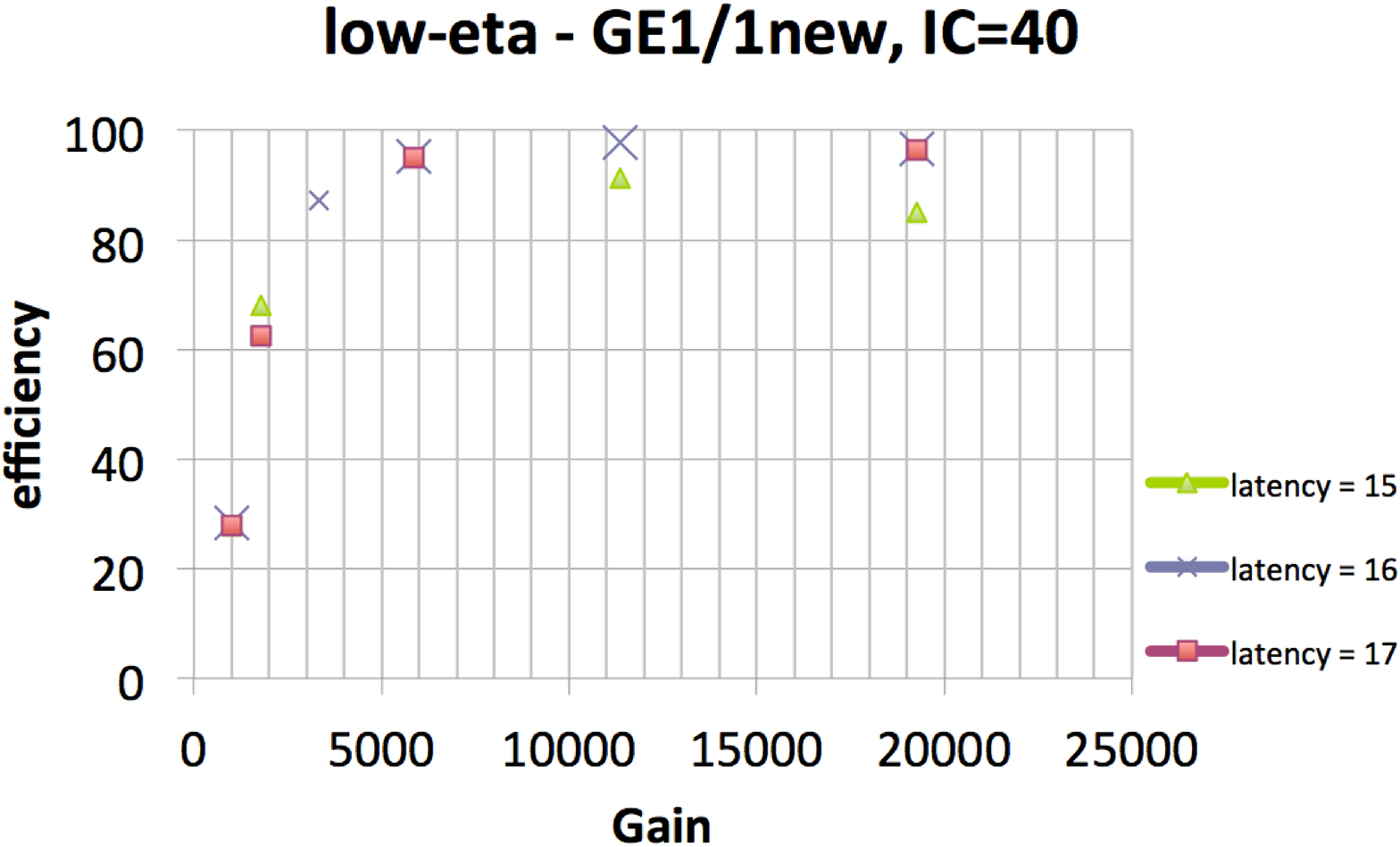}
  \caption{GE1/1$\_$II efficiency performance at high and low $\rm{\eta}$ ends. Efficiency reaches 96.5\% at gain 7000.} 
 \label{ge11_performance1}
\end{figure}

The GE1/1$\_$II was tested at the H8 and H4 muon beam lines at the SPS CERN in June, July, August, and September 2011.
Dedicated test beam measurements were performed at the SPS to study efficiency, space resolution, and timing performance using different gas mixtures and a strong magnetic field of 3T.
Several gas mixtures and different configurations were used in all the data-taking periods.
Based on the very low observed noise in the detector without beam, a VFAT2 threshold of 12 units\footnote{One VFAT threshold unit corresponds to a charge of $\rm{\approx0.08 fC}$ at the input channel comparator stage.} was set along with a comparator current set to $\rm{40\mu A}$. With these settings the noise was practically absent, when operating the detector in a gain range of $\rm{0-10^4}$. 
Fig.~\ref{ge11_performance1} shows a preliminary HV and latency scan with $\rm{Ar/CO_2/CF_4}$ (45:15:40).

\begin{figure}[H]
  \centering
  \includegraphics[width=2.8in]{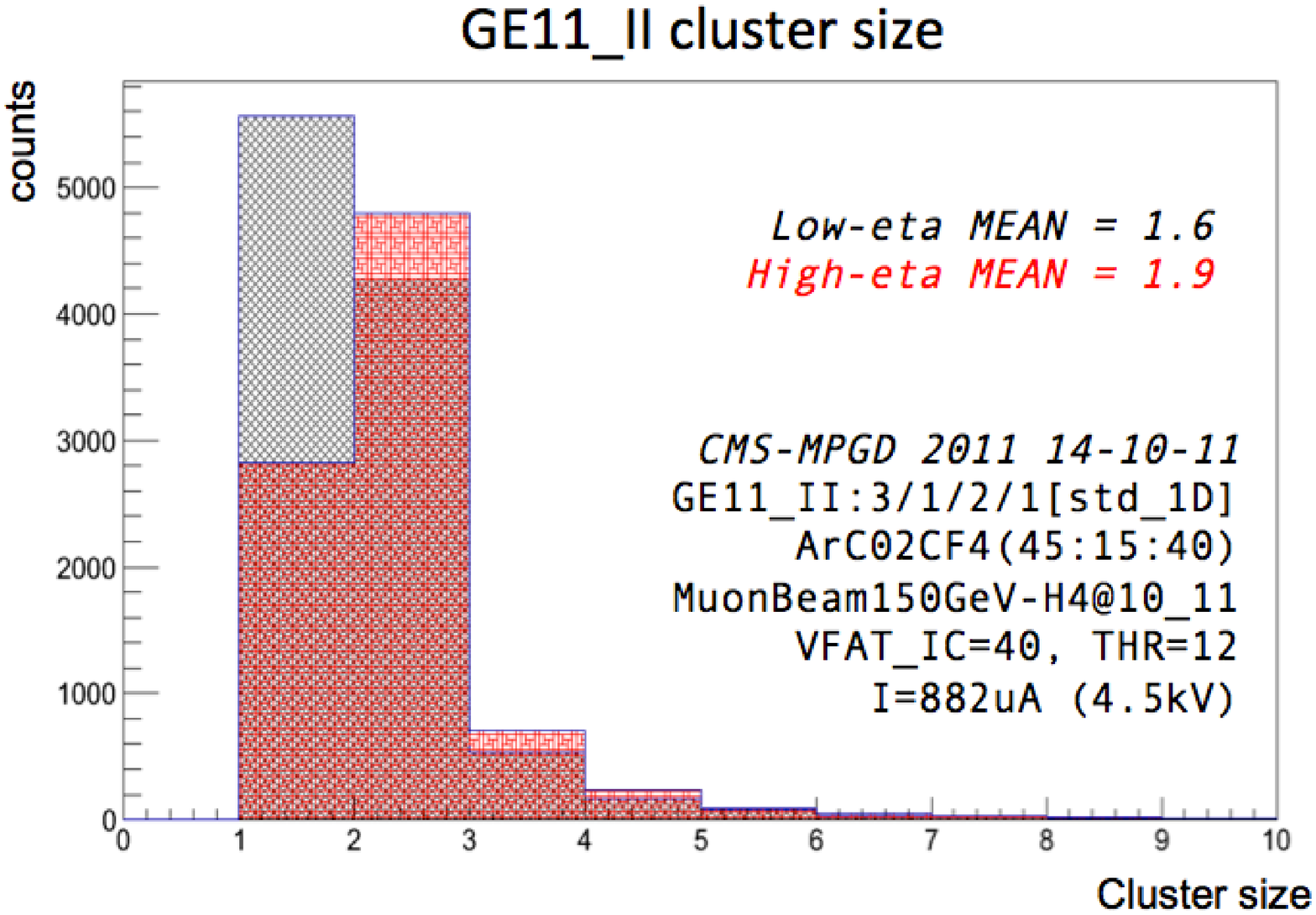}
  \includegraphics[width=2.7in]{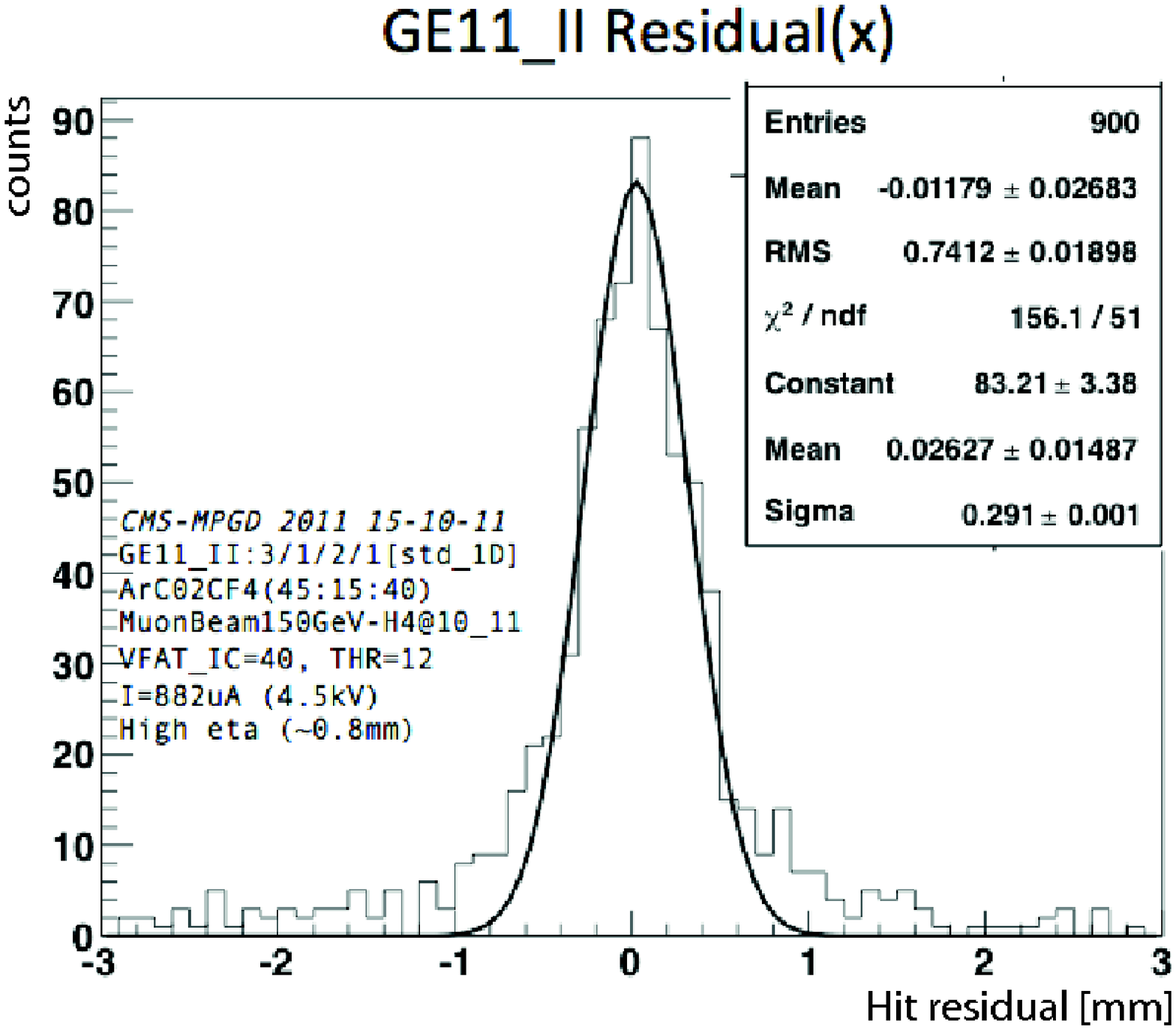}
  \caption{GE1/1$\_$II performance: space resolution and cluster size in the high and low $\rm{\eta}$.}
  \label{ge11_performance2}
\end{figure}

A remarkable result is the fact that the GE1/1$\_$II prototype reaches full efficiency with a gain of $\approx$7000: this is in agreement with previous results on small prototypes and indicates that the full-size detector is performing excellently; the operation at higher gain will ensure a very stable operation.
Fig.~\ref{ge11_performance2} shows the cluster size and achieved space resolution.

\begin{figure}[H]
  \centering
  \includegraphics[width=2.75in]{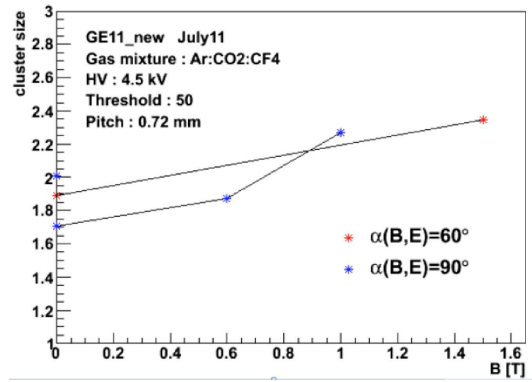}
  \caption{GE1/1$\_$II performance inside a strong magnetic field: cluster size.}
  \label{magnet_clusters}
\end{figure}

\begin{figure}
  \centering
  \includegraphics[width=2.7in]{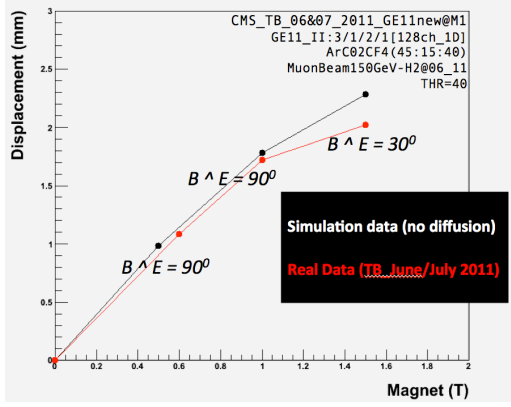}
  \caption{GE1/1$\_$II performance inside a strong magnetic field: strip cluster displacement due to the magnetic field.}
  \label{ge11_magnet_displ}
\end{figure}

In one campaign a strong magnetic field was used to validate the detector performance in an environment similar to the high-$\eta$ region of the CMS muon endcap. In  Figs.~\ref{magnet_clusters} and \ref{ge11_magnet_displ} measured strip cluster sizes and cluster displacements are shown. The cluster size does not appear to be affected by the magnetic field, while the signal induced on the strips is displaced due to the presence of the magnetic field. The measurement of this displacement is in good agreement with simulations performed with GARFIELD.

In August 2011 the full-scale prototype GE1/1$\_$II was tested again in H4 SPS beam line. This time the goal was to test the detector with the Scalable Readout System (SRS), provided by the RD51 Collaboration, and the APV25 chips. During those tests the capability to measure spatial resolution using full pulse height information, was shown. Two small $\rm{10 \times 10cm^2}$ GEM detectors (called $TR_5$ and $TR_1$) were used to reconstruct the tracks in the same manner as was previously done using the VFAT2 readout chip. Both these small detectors were running with $\rm{Ar/CO_2}$ while the GE1/1$\_$II was operated with $\rm{Ar/CO_2/CF_4}$. 
To minimize the impact of beam divergences on the measurement, tracks were selected to be from a $\rm{2 \times 2mm^2}$ spot in the center of a less divergent pion beam.

\begin{figure}[H]
  \centering
  \includegraphics[width=2.7in]{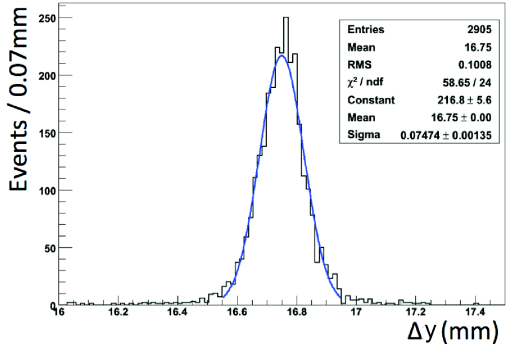}
  \caption{$\rm{\Delta y}$ distribution for $TR_5$ and $TR_1$.}
  \label{apv1}
\end{figure}

Assuming that both $TR_5$ and $TR_1$, which share the same construction, have the same spatial resolutions ($\rm{\sigma_{y5} = \sigma_{y1}}$) and that the beam divergence in y is negligible in the center, Fig.\ref{apv1} shows that we have for the width of the $\rm{\Delta y}$ distribution:

 \begin{equation}
   \sigma^2_{\Delta y}  = \sigma^2_{y5} + \sigma^2_{y1} = 2 \times \sigma^2_{y}
\end{equation}

 \begin{equation}
  \sigma_{y}  = \frac{\sigma_{\Delta y}}{\sqrt{2}} = 53 \mu m
\end{equation}

Hit positions in $x$ and $y$ are computed from the mean (or "center-of-gravity") of the corresponding strip cluster using analog pulse height information for each strip.
It is a reasonable assumption that $\rm{  \sigma_{x5} \approx 53 \mu m  }$ if $\rm{  \sigma_{y5} = 53 \mu m  }$ since in $TR_5$ strips in x and y have the same pitch. Consequently, an upper bound on the spatial resolution of the GE1/1$\_$II prototype can be established, in the chamber section with smallest strip pitch at the high-$\eta$ end when analog pulse height information is used:

\begin{equation}
 \label{ge11_upperbound}
  \sigma_{x_{GE11}}  \le  \sqrt{ \sigma^2_{\Delta x} - \sigma^2_{x5} } = 103 \mu m
\end{equation}

\begin{figure}[H]
  \centering
  \includegraphics[width=2.7in]{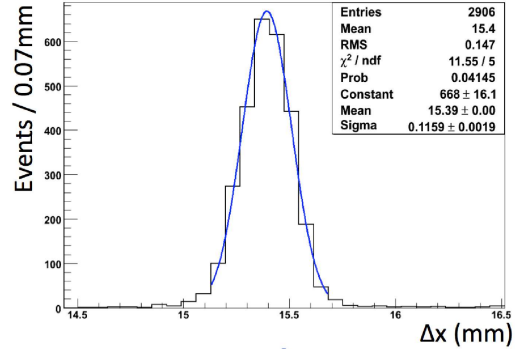}
  \caption{$\rm{\Delta x}$ distribution for $TR_5$ and GE1/1$\_$II using central tracks in a pion beam.}
  \label{apv2}
\end{figure}

Equ. (\ref{ge11_upperbound}) gives an upper bound value, as any remaining beam divergences in $x$ will still contribute to the width of the $\rm{\Delta x}$ distribution in Fig.~\ref{apv2}.

\section{Conclusion}
The GE1/1$\_$II detector was tested during 2011 in several test beams showing very positive results. The GEM collaboration (GEMs for CMS) made a significant improvement in the construction and the assembly of this second full-scale detector since it was challenging to build a large-area GEM with transfer and induction gap sizes as small as 1mm. Good detector performances on detection efficiency and spatial resolution were demonstrated and the detector behaved as expected when operated in a magnetic field. Further hardware development will be implemented in the future prototypes.

\section{Acknowledgment}
This work has partly been performed in the framework of the RD51 Collaboration.


\end{document}